\documentclass{jnmp03c}

\begin{document}
\setcounter{page}{50}

\renewcommand{\evenhead}{L A Khodarinova and I A Prikhodsky}
\renewcommand{\oddhead}{On Algebraic Integrability of
 the Deformed Elliptic Calogero--Moser Problem}

\thispagestyle{empty}

\FistPageHead{1}{\pageref{khodarinova-firstpage}--\pageref{khodarinova-lastpage}}{Letter}

\copyrightnote{2001}{L A Khodarinova and I A Prikhodsky}

\Name{On Algebraic Integrability of
 the Deformed Elliptic Calogero--Moser Problem}\label{khodarinova-firstpage}

\Author{L~A~KHODARINOVA~$^{\dag}$ and I~A~PRIKHODSKY~$^{\ddag}$}

\Adress{$^{\dag }$~Department of Mathematics, Statistics and Operational Research\\
~~The Nottingham Trent University, Burton Street, Nottingham NG1 4BU, UK\\
~~E-mail: Larisa.Khodarinova@ntu.ac.uk\\[2mm]
$^{\ddag}$~Institute of Mechanical Engineering, Russian Academy of Sciences\\
~~M. Haritonievsky, 4, Centre, Moscow, 101830, Russia}

\Date{Received July 13, 2000; Accepted September 15, 2000}

\begin{abstract}
\noindent
 Algebraic integrability of the elliptic Calogero--Moser quantum
problem related to the deformed root systems $\pbf{A_{2}(2)}$ is
proved. Explicit formulae for integrals are found.
\end{abstract}

\noindent
Following to \cite{khod:Kr} (see also~\cite{khod:ChV} and \cite{khod:VSCh}) we call a
Schr\"{o}dinger operator
\[
L=-\Delta +u\left( \vec x\,\right) ,\qquad \vec x \in R^{n},
\]
{\it integrable} if there exist $n$ commuting differential operators
$L_{1}=L,$ $L_{2},\ldots ,L_{n}$ with constant algebraically independent
highest symbols $P_{1}\left( \vec \xi \,\right) =\left(
\vec \xi \,\right) ^{2},$ $P_{2}\left( \vec \xi \,\right) ,\ldots ,
P_{n}\left( \vec \xi \,\right)$,
 and {\it algebraically integrable} if there exists at least one
more differential operator $L_{n+1},$ which commutes with the operators
$L_{i}$, $i=1,\ldots ,n,$ and whose highest symbol $P_{n+1}\left( \vec\xi \,\right) $
is also independent on $x$ and takes different
values on the solutions of the algebraic system
\begin{equation}
P_{i}\left( \vec \xi \, \right) =c_{i},\qquad i=1,\ldots ,n,
\label{khod:odin}
\end{equation}
for generic $c_{i}.$

The question how large is the class of the algebraically integrable
Schr\"{o}\-din\-ger operators is currently far from being understood, so any
new example of such an operator is of substantial interest.

The main result of this paper is the proof of the algebraic integrability of
the following Schr\"{o}dinger operator
\begin{equation}
L=-\frac{\partial ^{2}}{\partial x_{1}^{2}}-\frac{\partial ^{2}}{\partial
x_{2}^{2}}-m\frac{\partial ^{2}}{\partial x_{3}^{2}}+2(m+1)\left( m\wp
(x_{1}-x_{2})+\wp (x_{1}-x_{3})+\wp (x_{2}-x_{3})\right)  \label{khod:dva}
\end{equation}
when the parameter $m=2.$ Here $\wp $ is the classical Weierstrass
elliptic function satisfying the equation $\wp ^{\prime
}{}^{2}-4\wp ^{3}+g_{2}\wp +g_{3}=0.$ The operator
(\ref{khod:dva}) was introduced by Chalykh, Feigin and Veselov in
\cite{khod:VFC} (see also \cite{khod:CFV}) and is related to the
deformed root system $\pbf{A_{2}(m)}$ (see
\cite{khod:VFC},\cite{khod:CFV} for details).

When $m=1$ this is the well-known three-particle Calogero--Moser problem.
The usual integrability of this problem has been established by Calogero,
Marchioro and Ragnisco in~\cite{khod:CMR}, the algebraic integrability has been
proved in~\cite{khod:Kh} and in a more general case in~\cite{khod:BEG}.

The deformed Calogero--Moser\ system~(\ref{khod:dva}) in the
trigonometric and rational limits has been completely investigated
by Chalykh, Feigin and Veselov in~\cite{khod:VFC},\cite{khod:CFV},
where the algebraic integrability for the corresponding systems
has been proved for any $m.$ They have also conjectured that the
same is true in the elliptic case.

In this paper we prove this conjecture for $m=2.$ The first results in this
direction have been found in~\cite{khod:Kh}, where it was proved that
problem~(\ref{khod:dva}) is integrable. The corresponding integrals have the form
\begin{equation}
\ba{l}
\ds L_{1}=L=-\partial _{1}^{2}-\partial _{2}^{2}-m\partial
_{3}^{2}{}+2(m+1)\left( m\wp _{12}+\wp _{13}+\wp _{23}\right) ,
\vspace{2mm}\\
\ds L_{2}=\partial _{1}+\partial _{2}+\partial _{3},
\vspace{2mm}\\
\ds L_{3}=\partial _{1}\partial _{2}\partial _{3}+\left( \frac{1-m}{2}\right)
(\partial _{1}+\partial _{2})\partial _{3}^{2}+\left( \frac{1-m}{2}\right)
\left( \frac{1-2m}{3}\right) \partial _{3}^{3}
\vspace{2mm}\\
\ds \phantom{L_{3}=}+(m+1)\left( \wp _{23}\partial _{1}+\wp _{13}\partial _{2}\right)
+m(m+1)\wp _{12}\partial _{3}
\vspace{2mm}\\
\ds \phantom{L_{3}=}+\left( \frac{1-m}{2}\right) (m+1)\left( (\wp _{13}+\wp
_{23})\partial _{3}+\partial _{3}(\wp _{13}+\wp _{23})\right) ,
\ea
\label{khod:tri}
\end{equation}
where we have used the notations $\partial _{i}={\partial }/{\partial x_{i}}$,
$\wp _{ij}=\wp (x_{i}-x_{j}).$

It was also shown that the operator
\be\label{khod:chetyre}
\ba{l}
\ds L_{12} = (\partial _{1}-m\partial _{3})^{2}(\partial _{2}-m\partial
_{3})^{2}
\vspace{2mm}\\
\ds \phantom{L_{12}=}
+2(m+1)^{2}\wp _{23}(\partial _{1}-m\partial _{3})^{2}-2(m+1)^{2}\wp
_{13}(\partial _{2}-m\partial _{3})^{2}
\vspace{2mm}\\
\ds \phantom{L_{12}=}
+2m(m+1)\left( \wp _{12}-\wp _{13}-\wp _{23}\right) (\partial
_{1}-m\partial _{3})(\partial _{2}-m\partial _{3})
\vspace{2mm}\\
\ds \phantom{L_{12}=}
-m(m+1)\left( \wp _{12}^{\prime }+m\wp _{13}^{\prime }+3(m+1)\wp
_{23}^{\prime }\right) (\partial _{1}-m\partial _{3})
\vspace{2mm}\\
\ds \phantom{L_{12}=}
-m(m+1)\left( -\wp _{12}^{\prime }+m\wp _{23}^{\prime }+3(m+1)\wp
_{13}^{\prime }\right) (\partial _{2}-m\partial _{3})
\vspace{2mm}\\
\ds \phantom{L_{12}=}
-m(m+1)\wp _{12}^{\prime \prime }-3/2m^{2}(m+1)^{2}\wp _{13}^{\prime
\prime }-3/2m^{2}(m+1)^{2}\wp _{23}^{\prime \prime }
\vspace{2mm}\\
\ds \phantom{L_{12}=}
+m^{2}(m+1)^{2}\left( \wp _{12}^{2}+\wp _{13}^{2}+\wp _{23}^{2}\right)
+2m(m+1)^{2}\left( \wp _{12}\wp _{13}+\wp _{12}\wp _{23}\right)
\vspace{2mm}\\
\ds \phantom{L_{12}=}
{}+2(m+1)^{2}(2m^{2}+3m+2)\wp _{13}\wp _{23}
\ea
\ee
commutes with $L_{1}$, $L_{2}$, $L_{3}$ and therefore is an additional
integral of the problem~(\ref{khod:dva}). Unfortunately, this is not enough for
algebraic integrability of the problem~(\ref{khod:dva}) since the highest symbol
of $L_{12}$ is invariant under permutation of $\xi _{1}$ and $\xi _{2}$ and
therefore takes the same values on some of the solutions of the
corresponding system~(\ref{khod:odin}).

In this paper we present an explicit formula of one more integral for the
system~(\ref{khod:dva}) related to the deformed root systems $\pbf{A_{2}(2).}$
This integral together with the previous integrals guarantees the algebraic
integrability of the system (\ref{khod:dva}) in case of $m=2.$

\medskip

\noindent {\bf Theorem.} {\it The system (\ref{khod:dva}) with
$m=2$ besides the quantum integrals given by~(\ref{khod:tri})
and~(\ref{khod:chetyre}) has also the following integral
$L_{13}=I+I^{\ast }$,  where
\[
\ba{l}
\ds I = \frac{1}{2}(\partial _{1}-\partial _{2})^{4}(\partial _{1}-2\partial
_{3})^{2}
-9\wp _{13}(\partial _{1}-\partial _{2})^{4}-24\wp _{12}(\partial
_{1}-\partial _{2})^{2}(\partial _{1}-2\partial _{3})^{2}
\ea
\]
\be\label{khod:5} \ba{l} \ds \phantom{I=} -6(\wp _{12}+\wp
_{13}-\wp _{23})(\partial _{1}-\partial _{2})^{3}(\partial
_{1}-2\partial _{3}) +\Bigl( 414\wp _{12}\wp _{13}+18\wp _{12}\wp
_{23}+18\wp _{13}\wp _{23}
\vspace{2mm}\\
\ds \left.\phantom{I=} +72\wp _{12}^{2}+108\wp _{13}^{2}+36\wp
_{23}^{2}-\frac{201}{2}g_{2} \right) (\partial _{1}-\partial
_{2})^{2} +\Bigl(144\wp _{12}\wp _{13}-144\wp _{12}\wp _{23}
\vspace{2mm}\\
\ds \phantom{I=}
+432\wp _{12}^{2}+18\wp _{13}^{2}-18\wp _{23}^{2}+33g_{2}
\Bigr) (\partial _{1}-\partial _{2})(\partial
_{1}-2\partial _{3})
\vspace{2mm}\\
\ds \phantom{I=}
+\left(288\wp _{12}^{2}-69g_{2}\right)(\partial _{1}-2\partial _{3})^{2}-369\wp
_{12}^{\prime }\wp _{13}^{\prime }+288\wp _{12}^{\prime }\wp _{23}^{\prime
}+18\wp _{13}^{\prime }\wp _{23}^{\prime }
\vspace{2mm}\\
\ds \phantom{I=}
-5760\wp _{12}^{3}-648\wp _{13}^{3}-288\wp _{23}^{3}-\wp _{12}^{2}(3834\wp
_{13}+1350\wp _{23})
\vspace{2mm}\\
\ds \phantom{I=}
+\wp _{13}^{2}(594\wp _{12}-594\wp _{23})-\wp _{23}^{2}(648\wp _{12}-324\wp
_{13})
\vspace{2mm}\\
\ds \phantom{I=}
+g_{2}\left( \frac{5085}{2}\wp _{12}+\frac{2061}{2}\wp _{13}+990\wp
_{23}\right)
\ea \hspace{-15mm}
\ee
and $I^{\ast }$ is the operator adjoined to $I$.

The integral $L_{4}=L_{13}+\frac{1}{2}L_{23}$, where $L_{23}$
is given by the same formula~(\ref{khod:5}) after permutation of $x_{1}$
and $x_{2}$ and $\partial _{1}$ and $\partial _{2}$,
is an additional integral which together with $L_{1}$,
$L_{2}$, $L_{3}$ guarantees the algebraic integrability.}

\medskip

Let us first comment on how this new integral $L_{13}$ has been
found. The highest symbol has been borrowed from the trigonometric
case~\cite{khod:VFC},\cite{khod:CFV}. The commutativity relation
between this integral and the Hamiltonian~$L_{1}$ imposes a very
complicated overdetermined system of relations on the coefficients
of the integral. We have resolved these relations combining the
direct analysis with the use of a computer. The addition theorem
and the differential equations for the elliptic $\wp$-function
play the essential role in these calculations. The fact that this
overdetermined system has a solution seems to be remarkable.

It is obvious from the explicit formula that $L_{13}$ commutes with $L_{2}.$
The commutativity of $L_{13}$ and $L_{3}$ has been checked with the help of
a computer. We have used a special program, which has been created for this
purpose, and the same technical tricks as in our previous paper~\cite{khod:KhP}.
The commutativity of the operators $L_{1}$, $L_{2}$, $L_{3}$ has been proved
in~\cite{khod:Kh}.

It is easy to check that the highest symbol of $L_{4}$ takes different
values on the solutions of the corresponding system~(\ref{khod:odin}). This
completes the proof of the algebraic integrability.

\medskip

\noindent
{\bf Remark.} We should mention that according to Krichever's general result
(see \cite{khod:Kr}) integrals $L_{1}$, $L_{2}$, $L_{3}$, $L_{4}$ satisfy certain
algebraic relations (spectral relations). In \cite{khod:KhP} we have found
explicitly these relations in the non-deformed case $m=1.$ In the deformed
case it seems to be a much more involved problem.

\medskip

\noindent
{\bf Acknowledgements.}
The authors are grateful to O~A~Chalykh and A~P~Veselov, who attracted
our attention to this problem.

\label{khodarinova-lastpage}

\end{document}